\documentclass[aps,prl,reprint,superscriptaddress]{revtex4-2}

\usepackage{graphicx}
\usepackage{dcolumn}
\usepackage{bm}

\usepackage[utf8]{inputenc}
\usepackage[english]{babel}
\usepackage{amsmath}
\usepackage{amsfonts}
\usepackage{csquotes}
\usepackage{indentfirst}
\usepackage{subfigure}

\usepackage[T1]{fontenc}
\usepackage{mathptmx}

\usepackage{enumitem}
\graphicspath{ {./figures/} }

\begin{document}


\title{Accurate Estimate of the Joint Density of States via Flat Scan Sampling}

\author{J. C. In\'{a}cio}
\affiliation{Departamento de Física and CICECO - Aveiro Institute of Materials, 3810-193 Aveiro, Portugal}

\author{A. L. Ferreira}
\affiliation{Departamento de Física and i3N, 3810-193 Aveiro, Portugal}

\author{J. S. Amaral}
\email{jamaral@ua.pt}
\affiliation{Departamento de Física and CICECO - Aveiro Institute of Materials, 3810-193 Aveiro, Portugal}

\date{\today}

\begin{abstract}

    
    A Monte Carlo method to estimate the Joint Density of States \(g(E,M)\) of the Ising and Ising-like models is presented. The method is applied to the well-known 2D Ising model, and is shown to be accurate, efficient, and embarrassingly parallel. The method presented offers major improvements over existing approaches. Furthermore, we obtain \(g(E,M)\) estimates for the spin S Ising model, with the spin number \(S = \{1/2, 1, 3/2, 2\}\), thus showing that the algorithm can handle larger and more complex \((E, M)\) phase spaces.

\end{abstract}

\maketitle


    Computer simulations, nowadays, play a central role on the study of complex systems, phase transitions and critical phenomena in statistical mechanics and condensed matter physics \cite{landau_binder}. Models such as the Ising model, the Heisenberg model, the Potts model and spin glass models describe interesting properties of such systems that can be grasped through numerical simulation based on the Monte Carlo method. Exact solutions to such systems are extremely difficult or impossible to attain and hence the need for computer simulations.
    The go-to methods to simulate such systems are \textit{Markov Chain Monte Carlo} methods. These methods exploit the fact that, in the equilibrium regime of a Markov chain, we can generate a sequence of configurations based on some probability distribution. Usually, this probability is set to the Boltzmann distribution, \(P(E) = e^{-\beta E}/Z\), with \(\beta = 1/(k_B T)\). This makes every simulation temperature (and applied fields or magnetic anisotropy) dependent. So we are required to run a simulation for every temperature value if we want to predict how the system's thermodynamic properties change with temperature. The most well-known method in this class is the \textit{Metropolis method}, introduced in 1953 by Metropolis et al. \cite{Metropolis1953}.
    
    To avoid repeated simulations for each temperature value, we can estimate the Density of States \(g(E)\) (DoS), as it contains all of the physical information about the system. It is defined as the number of states of the system with a given value of energy, \(E\). If we are studying magnetic systems, it would be more interesting to look at the \textit{Joint Density of States} (JDoS), defined as \(g(E, M)\), a function of both energy \(E\) and magnetization \(M\). With the JDoS, we can compute the partition function 
    \begin{equation*}
        Z(T) = \sum_{i, q} g(E_i, M_q) e^{- \beta H(E_i, M_q)},
    \end{equation*}
    where \(H\) is the system Hamiltonian \(H(E, M) = E - Mh\), with \(h\) as an external magnetic field (here the index \(i\) runs over all possible values of energy and \(q\) over all of the possible values of magnetization). From this we are able to determine the free energy as a function of the temperature \(T\) and magnetization \(M\), \(F(T, M)\), and thus all of the thermodynamic properties of the system.
    
    
    
    More recently a new class of Monte Carlo methods has emerged, the \textit{Reweighting methods} \cite{landau_binder}. Some of these methods are characterized by a random walk through the phase space that produces a flat histogram. This corresponds to sampling possible system configurations with a probability proportional to the inverse of the DoS (or JDoS) value at that particular state. These methods came to solve the problem of temperature dependent simulations and the need to run many simulations to precisely characterize thermodynamic behaviour near phase transitions. Additionally, by not being constrained by the Boltzmann distribution, these methods avoid the Free Energy landscapes and barriers that arise by sampling from the Boltzmann distribution.
    
    The main difficulty of a random walk through the phase space is that the walker would be stuck on the most probable macrostates thus proving the random walk ineffective. The \textit{Wang-Landau method} (WL), proposed in 2001 by Fugao Wang and David P. Landau \cite{WL_org}, bypasses this by performing a random walk with a probability proportional to the inverse of the density of states \(1/g(E, M)\). Then, we ensure that every macrostate has the same probability of being visited during the random walk. Since \(g(E, M)\) is unknown before the simulation we have to start with a guess (usually \(g(E, M) = 1\)) and update it, through a modification factor \(f\), each step of the random walk. We stop the random walk when have achieved a flat histogram in \((E, M)\) phase space, with a margin of \(p\%\). Since the first steps in the random walk are highly biased, we accumulate statistical errors in the estimation of \(g(E, M)\). These errors can be reduced, but never eliminated, by letting \(f_{final} \rightarrow 1\) and \(p \rightarrow 100\) \cite{WL_org, WL_S_parm}.
    
    The parallel efficiency and ease of implementation of a computational method is also an important aspect when trying to simulate large systems. WL's parallelization scheme is not trivial, since you have to divide the phase space between different walkers and have them communicate at the boundaries. This causes some parallel overhead which dampens performance \cite{WL_parallel}.
    
    
    A few modifications to the original WL algorithm have been proposed, namingly the modifications by C. Zhou et al. \cite{WL_S_parm}, and Belardinelli et al. \cite{WL_t_method}. The first proposed an introduction of an additional parameter that reduces correlations between successive updates to \(g(E, M)\); the later, although not originally presented for the estimation of \(g(E, M)\), introduced a new way of updating \(g(E)\). Both improvements saw a reduction in the statistical errors present in WL, however both increase computing time. 
    
    
    In this work, we introduce an efficient and highly parallelizable Monte Carlo method that offers some benefits over existing approaches. We named it \textit{Flat Scan Sampling}, FSS for short. Unlike WL, where we perform a random walk through the whole phase space, our approach is to sample an energy slice of \((E, M)\) phase space, where \(M=M_q\) and \(g(E)\) is known, and construct an energy histogram at \(M=M_{q+1}\) that holds information about the DoS at \(M_{q+1}\). Thus sampling the phase space sequentially, magnetization by magnetization. We sample the slices \((E, M_q)\) of the phase space through a random walk with acceptance ratio of \(1/g(E,M)\), to efficiently sample configurations at \(M_q\), over all possible energy values observed. Since \(g(E, M_q)\) is known beforehand, the random walk will obey the desired detailed balance. Through the random walk at the energy space of \(M_q\), we are able to evaluate \(h(E_{a}, E_{b})\), an energy histogram, where \(E_a\) and \(E_b\) are the initial (at \(M_q\)) and final (at \(M_{q + 1}\)) energy values, respectively. Performing this sequential sweep through the phase space, an accurate estimate of \(g(E,M_{q+1})\) is obtained.
    
    The process through which we construct the energy histogram is called \textit{scan}. In this procedure, we compute the energy of a new configuration, generated by sequentially flipping the spins that lead from \(M_q\) to \(M_{q+1}\). This leads to the discovery of the possible energy values at the next magnetization. After computing the energy, we update \(h(E_a, E_b)\) and revert the spin back to its original state. This is repeated for every up spin in the original configuration.
    
    The validity of the FSS method can be shown explicitly through the formalism of the transition matrix. Consider a random walk through a slice of the \((E, M)\) space, where \(M=M_q\). Through the scanning steps, we evaluate the energy histogram \(h(E_a, E_b)\), with \(E_a\) and \(E_b\) as the initial and final energy values, respectively. Defining the infinite temperature \textit{Transition Matrix} (TM) \(W(E_a, E_b)\) \cite{TMMC, deoliveira1996broad} as the fraction of possible spin-flip transitions that lead from states with energy \(E_a\) at magnetization \(M_q\) to states with energy \(E_b\) at magnetization \(M_{q+1}\). Since we have \(\sum_{E_b} W(E_a, E_b) = 1\) and
    \begin{equation*}
        h(E_a, E_b) = \frac{W(E_a, E_b)}{g(E_a)},
    \end{equation*}
    we can obtain \(\sum_{E_b} h(E_a, E_b) = 1 / g(E_a)\) and conclude that the normalized histogram is an estimator of the TM
    \begin{equation*}
        h^{(n)} (E_a, E_b) = \frac{h(E_a, E_b)}{\sum_{E_b} h(E_a, E_b)} = W(E_a, E_b).
    \end{equation*}
    Since \(W(E_a, E_b) g(E_a) = W(E_b, E_a) g(E_b)\), we obtain
    \begin{equation*}
        g(E_b) = \sum_{E_a} W(E_a, E_b) g(E_a),
    \end{equation*}
    by summing over \(E_a\). This means that from the histogram of initial and final energy values of possible spin-flips of states generated according to the probability \(1 / g(E_a)\) at \(M_q\), we can determine \(g(E_b)\) at \(M_{q+1}\) by using
    \begin{equation} \label{eq:jdos_fss}
        g(E_b) = \sum_{E_a} h^{(n)} (E_a, E_b) g(E_a).
    \end{equation}


    

    The energy histogram \(h(E_a, E_b)\) is estimated through the sampling of a desired number of configurations with a given energy value, \(E_a\). The higher this number (here referred to as \(REP\), and defined as the number of samples per energy point), the more accurate the DoS estimation at \(M_{q+1}\) will be. If the number of samples is very small, the simulation might not be able to converge, due to large errors in the estimated DoS. 
    We can further improve the performance of FSS by the introduction of a new parameter that we call \(skip\) (similar to the parameter introduced by Zhou et al. \cite{WL_S_parm}). This parameter dictates the number of steps in the random walk before the algorithm is allowed to scan a new configuration. This will reduce the statistical correlation between scanned configurations lowering the overall error. 

    This way, the algorithm goes as follows:
    \begin{enumerate}[noitemsep,nolistsep]
        \item Choose a magnetization value \(M_q\) where \(g(E, M_q)\) is already known. Usually, we start with all spins up;
        \item Generate a random configuration with that magnetization and compute its energy \(E_a\);
        \item Perform two spin flips to generate a new configuration in the same magnetization and compute its energy \(E_b\);
        \item Accept the new configuration with the probability given by 
        
        \begin{equation*} 
            P(E_a \rightarrow E_b) = \min \left\{1, \frac{g(E_a)}{g(E_b)}\right\};
        \end{equation*}
        
        \item If the given configuration has been sampled more than \(REP\) times or it has been less than \(skip\) steps since the last scan, continue. Else, scan the accepted configuration \((E_a, M_q)\), i.e., for every spin up in the configuration, perform:
        \begin{enumerate}[noitemsep,nolistsep]
            \item Flip the spin down and compute the new energy of the configuration \(E_b\);
            \item Update the histogram \(h(E_a, E_b)\);
            \item Revert the spin flip.
        \end{enumerate}
        \item If the number of scanned configurations per value of energy is equal to \(REP\), then stop the simulation and compute the DoS at the next magnetization value \(M_{q+1}\), using Equation \eqref{eq:jdos_fss}.
        
    \end{enumerate}
    
    Parallelization for this method can be achieved in a simple way. Given that the random walk in the energy space only depends on the estimation of the previous iteration, we can have multiple independent walkers walkers sampling their own energy histogram \(h_i\), \(i=1,\ldots,n\), with \(n\) as the number of walkers, at the same magnetization value \(M_q\). When all of the walkers end their random walk, we may compute \(g(E)\) at the next magnetization \(M_{q+1}\), by using the sum of all of the sampled energy histograms \(h = \sum_i h_i\). Seeing that we are just performing \(n\) simultaneous random walks, the performance of this parallelization scheme should have a scaling, in computational time, close to linear, as we increase the number of walkers \(n\). This is reinforced by the fact that during the random walk itself, there is no need for communication between walkers. This scheme allows for two interesting cases: \((1)\) simulating small or medium sized systems by having \(n\) simulations running in parallel with the intended value of \(REP\), generating results with a larger value of \(REP\); \((2)\) simulating a larger system by distributing the intended \(REP\) value by the \(n\) walkers, \(REP_{\text{walker}} = REP / n\). On case \((2)\), we have observed that when the \(REP_{\text{walker}}\) is low, the walker might not be able to sample all of the phase space correctly, missing some energy values. One way to avoid this is to discard the histograms from walkers that did not discover all of the energy space. 

\begin{figure}[t]
        \centering
        
        \subfigure[]{%
        \includegraphics{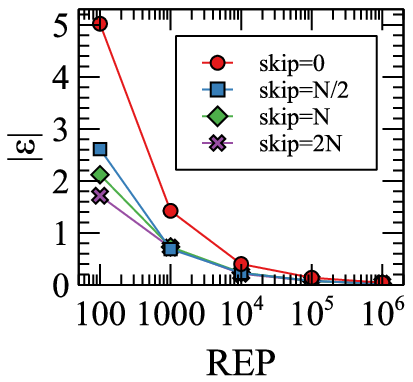}
        \label{subfig:abs_error_fss}}
        \subfigure[]{%
        \includegraphics{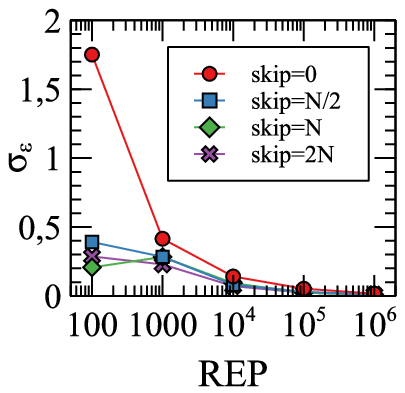}
        \label{subfig:std_error_fss}}
        
        \caption{Here \(N\) represents the number of spins in the system, in this case \(N=16\). The data was averaged over 100 independent simulations. (a) Absolute error in FSS of the JDoS, of a \(4 \times 4\) spin\(-1/2\) Ising lattice, versus the parameter \(REP\), for increasing values of \(skip\). (b) Standard deviation of FSS error of the JDoS, of a \(4 \times 4\) spin\(-1/2\) Ising lattice, versus the parameter \(REP\), for increasing values of \(skip\).}
        \label{fig:fss_val}
    \end{figure}

    The convergence and accuracy of our algorithm may be tested on a system composed by 16 Ising spins arranged on a square lattice. Since we know the exact JDoS for this system, through manual sampling of all possible configurations, it is possible to compare the estimated solution by the algorithm against the exact solution. We start by defining the mean relative error \(\epsilon = \mathbb{E} [\epsilon(g)] = \mathbb{E}[(g - g_{\text{exact}}) / g_{\text{exact}}]\) and the standard deviation of the error as \(\sigma_{\epsilon} = \sqrt{\mathbb{E} [(\epsilon(g) - \epsilon)^2]}\), where \(g_{\text{exact}}\) is the exact solution for the JDoS of an \(L=4\) simple squared Ising system. Here we treat the estimation of the JDoS by FSS, \(g\), as random variable. Simulations of such system were performed 100 times, with different \(REP\) and \(skip\) parameters. In Figure \ref{fig:fss_val}, the mean absolute error, Figure \ref{subfig:abs_error_fss}, and the standard deviation of the error, Figure \ref{subfig:std_error_fss}, of FSS computations can be seen. Independently of the \(skip\) value used, as we increase the number of samples, the absolute error of FSS converges linearly to zero. The same can be observed with the standard deviation of the error, meaning that our algorithm is both accurate and precise as we increase the parameter \(REP\), Figure \ref{fig:fss_val}. We can thus easily control the accuracy of our simulated results through the change of \(REP\).

    \begin{figure}[t]
        \centering
        \includegraphics{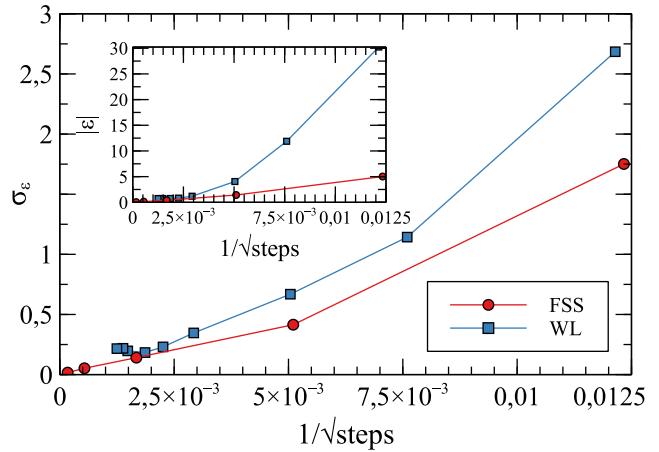}
        \caption{Standard deviation of the error of the JDoS, of a \(4 \times 4\) spin\(-1/2\) Ising lattice, computed by FSS and WL as a function of the inverse square root of the number of random walk steps, with \(skip=0\) and \(p=90\), respectively. Data was averaged for 100 simulations. Inset: Absolute error of FSS and WL as a function of the inverse squared root of the number of random walk steps, for \(skip=0\) and \(p=90\).}
        \label{fig:std_fss_wl}
    \end{figure}
    
    Shifting our focus to the effects of the \(skip\) parameter, we observe that it changes the precision of the computations by a small margin when the number of samples is high. However, for lower \(REP\) values, as the \(skip\) increases, the error diminishes. This is due to the reduction of statistical correlations between the limited number of samples, induced by the increase of \(skip\). A good rule of thumb is to set \(skip\) to a value between \(N/4\) and \(N/2\), where \(N\) is the number of spins on the system. 
    
    
    The standard deviation of well behaved Monte Carlo methods has a linear relationship with the inverse of the squared root of the number of samples used to estimate the solution. Thus, 
    \begin{equation} \label{eq:wb_mc}
        \sigma_{\epsilon} \propto \frac{1}{\sqrt{\text{steps}}}
    \end{equation}
    holds. As the right hand side goes to zero, one would expect that the standard deviation would also vanish. In Figure \ref{fig:std_fss_wl}, the standard deviation of the error for FSS and WL as a function of the inverse squared root of the steps can be observed. FSS's standard deviation follows the linear relation described in Equation \eqref{eq:wb_mc}. Furthermore, extrapolating the standard deviation for infinite steps would yield \(\sigma_{\epsilon}(\infty) \approx 0\). WL's standard deviation plateaus as the number of steps increases hence does not converge to zero. Extrapolating the standard deviation for infinite steps, we have \(\sigma_{\epsilon} (\infty) \approx 0.20\).

    \begin{figure*}[t]
        \centering
        \includegraphics[width=\textwidth]{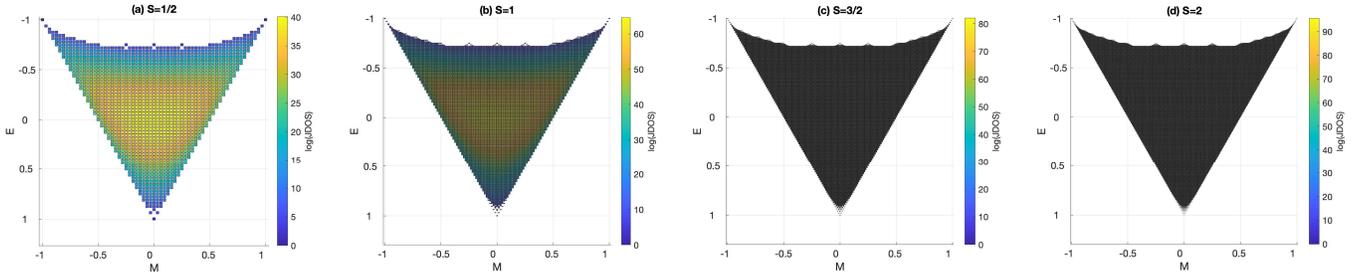}
        \caption{Logarithm of the JDoS for a spin S Ising system with 64 spins arranged on a square lattice estimated by the FSS method. From left to right: (a) S \(=1/2\) (b) S \(=1\) (c) S \(=3/2\) (d) S \(=2\)}
        \label{fig:spin_s_jdos}
    \end{figure*}

    \begin{figure}[t]
        \centering
        \includegraphics{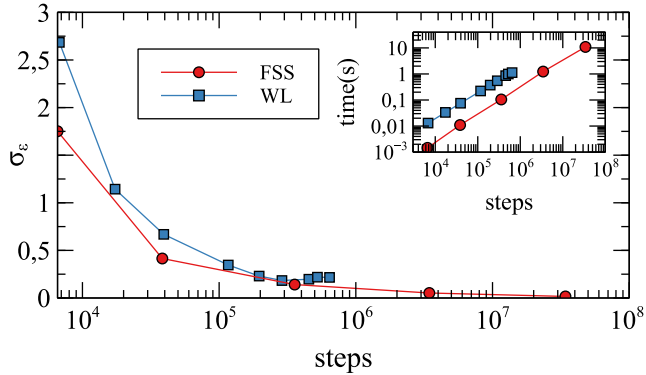}
        \caption{Standard deviation of the error of the JDoS of FSS and WL versus the number of steps in the random walk, for \(skip=0\) and \(p=90\), respectively. Inset: time of the computation, in seconds, as a function the random walk steps. Averages of \(100\) simulations were taken in order to reduce statistical noise.}
        \label{fig:std_wt_fss_wl}
    \end{figure}
    
    The efficiency of a computational method is of utmost importance. By efficiency, we mean the precision in the solution as a function of the steps taken to compute that solution. A more efficient method will have a higher precision estimation in less total steps. In Figure \ref{fig:std_wt_fss_wl}, the standard deviation of the error as a function of the number of steps is presented. We see that FSS has the same or equal precision for the same number of steps, compared to WL. However, FSS is able to lower its standard deviation given enough steps, while WL shows a plateau. Thus, FSS is shown to be more efficient than WL. 
    
    Another noteworthy point is the computational time per step. This will, of course, depend on implementation details, computer architecture, and many other variables. On the inset of Figure \ref{fig:std_wt_fss_wl}, we can observe the computation time in seconds as a function on steps. Here the absolute time is not important, but that WL and FSS scale the same way with regards to steps, but WL has an intrinsic time which makes the time per step higher. This emphasizes again that FSS is a more efficient method.
    
    \begin{table}[]
    \centering
    \caption{Wall time in seconds for FSS computations of 2D systems with \(L=\{4, 8, 16\}\). The benchmark was performed on an i7-4790 @ 3.6 GHz. The \(REP\) and \(skip\) values used were \(10^5\) and \(N\) (number of particles), respectively.}
    \begin{tabular}{c|c}
        \(L\) & Wall Time (s) \\ \hline
        \(4\) & 0.69 \\
        \(8\) & 126.56 \\
        \(16\) & 28638.31 
    \end{tabular}
    \label{tab:wall_time_fss}
    \end{table}
    
    Although not presented here, we have computed results for larger 2D systems with \(L=8\) and \(L=16\). Table \ref{tab:wall_time_fss}, shows the wall time, in seconds, for our implementation of FSS, for those systems. All of the results and codes used in this article can be found on the following GitHub page: \url{https://github.com/jgci2000/monte-carlo-ising} (to be made public). Here you can also find an implementation of FSS and WL in C++, ready to be used.
    
    
    We now apply the FSS algorithm to the more complex spin S Ising model. In this model, instead of considering spin\(-1/2\) particles, we have a lattice of spin\(-S\) particles. This is a quite more demanding scenario compared to the spin\(-1/2\), as the number of possible configurations and the size of the \((E, M)\) phase space drastically increase. Each spin is allowed \(2S+1\) possible states, meaning that each particle can have spin values in the \(z\) direction in the range of \(S_z \in \{-S , -S+1, \ldots, S-1, S\}\).
    
    
    Applying the FSS method to the Ising spin S model does not introduce substantial algorithm changes. While performing the random walk at a given $M_q$, the JDoS contribution to all $M$ values that can be reached by changing a single spin site state are accumulated throughout. The allowed values are the spin values that increase or decrease the magnetization of the system, depending on the direction of sampling. So, while for the spin 1/2 model the scan step contributes only to the DoS estimate at $M_{q+1}$, for the case of the spin 1 model, both DoS estimates at $M_{q+1}$ and $M_{q+2}$ are contributed to. The higher the spin value, more DoS contributions are accumulated at a single random walk at a given $M$ value. In a Transition Matrix perspective, this accumulation accounts for the extra non-zero elements of the Transition Matrix that result from increasing the model spin value. 
    
    In Figure \ref{fig:spin_s_jdos}, four JDoS computations, with the FSS method, for a system with 64 spins on a square lattice can be observed, for different numbers of the total spin \(S\). From left to right, \(S=1/2\), \(S=1\), \(S=3/2\) and \(S=2\). As shown, the size of the available phase space increases very rapidly with the spin degrees of freedom. For the \(S=1/2\) case, we have a \(65 \times 65\) phase space, whilst with \(S=2\), we will have a phase space with dimension \(1025 \times 257\). In such manner, estimating \(g(E, M)\) for the spin S model is considerably more demanding compared to the spin \(1/2\) case. 
    We have thus shown that the FSS method can accurately and efficiently estimate \(g(E, M)\) of increasingly complex discrete spin models. Furthermore one could explore the thermodynamic properties of complex spin models and their dependence on external parameters such as \(T\) and applied field.
    
    
    
    The method presented here, Flat Scan Sampling, is a novel computational method for an efficient and accurate estimation of the Joint Density of States of localized discrete spin systems. The algorithm samples the \((E, M)\) phase space by slices of magnetization and constructs an energy histogram at every iteration. This histogram was shown to be proportional to the Joint Density of States at the next magnetization, Equation \ref{eq:jdos_fss}. This method is embarrassingly parallel, with a large headroom for performance gains. After being validated and compared to the Wang-Landau sampling applied to the Ising model, we applied Flat Scan Sampling to the more general Ising spin S model. The method has shown it's capabilities of sampling \((E, M)\) phase spaces and estimating the Joint Density of States of more complex models.
    
    This work was developed within the scope of the project CICECO-Aveiro Institute of Materials, UIDB/50011/2020 and UIDP/50011/2020, financed by national funds through the FCT/MEC. JSA acknowledges FCT IF/01089/2015 grant. A. L. Ferreira acknowledges support from i3N projects Ref. Nos. UIDB/50025/2020, UIDP/50025/2020 and LA/P/0037/2020 financed by national funds through the FCT/MEC.
    
    \bibliography{FSS}

\end{document}